\documentclass[preprint,review]{elsarticle}

\usepackage[english]{babel}
\usepackage{color, graphicx}
\usepackage{soul}
\usepackage{float}
\usepackage{caption, subcaption}
\usepackage{amssymb}
\usepackage{amsmath}
\usepackage{xcolor}
\usepackage{booktabs}
\graphicspath{{figures/}}
\usepackage{epstopdf}
\usepackage{dsfont}
\usepackage{multirow}
\usepackage{color,soul}
\usepackage[utf8]{inputenc}
\usepackage{enumitem}
\usepackage [autostyle, english = american]{csquotes}
\MakeOuterQuote{"}

\biboptions{numbers,sort&compress}
\usepackage{textcomp}
\usepackage{lineno}
\usepackage{rotating}

\begin{document}
\begin{frontmatter}

\title{Full-scale validation of CFD simulations of buoyancy-driven ventilation in a three-story office building}
 
\author[y2e2]{Chen Chen\corref{cor1}}
\ead{chenc2@stanford.edu}
\cortext[cor1]{Corresponding author}

\author[y2e2]{Catherine Gorl\'e}
\ead{gorle@stanford.edu}

\address[y2e2]{Stanford University, Y2E2 Building, 473 Via Ortega, Stanford, CA, 94305}



\begin{abstract}
Computational fluid dynamics (CFD) is frequently used to support the design of naturally ventilated buildings; however, the model accuracy should be thoroughly assessed, ideally through validation with full-scale measurements. The present study aims to (1) validate transient CFD simulations with uncertainty quantification (UQ) for buoyancy-driven natural ventilation against full-scale experiments in an operational atrium building, and (2) quantify the sensitivity of the CFD results to the thermal boundary conditions.  The UQ and sensitivity analysis consider uncertainties in the initial and boundary conditions for the temperatures. Considering the volume-averaged air temperature on each floor, the predictions and measurements agree well with discrepancies less than 0.3 \textdegree{}C. When considering the temperature averaged over smaller zones on each floor, two trends can be observed. First, in zones not adjacent to windows, the discrepancies between the CFD and measurements can be explained by uncertainty in the boundary conditions and the measurements. Second, in zones adjacent to windows, higher discrepancies are observed due to oscillations in the inflow jets just downstream of the windows, and due to geometrical simplifications in the CFD model. The sensitivity analysis demonstrates that the boundary conditions for the thermal mass surface temperature and the outdoor temperature have a dominant effect on the indoor air temperature predictions, with their relative importance varying as function of proximity to the windows.
\end{abstract}


\end{frontmatter}

\section{Introduction}

Around 40\% of the total US energy consumption is in residential and commercial buildings, and a significant fraction ($\sim$45\%) of the energy consumption is for heating, ventilation, and air-conditioning~\cite{USEnergy2018}. Globally, increased urbanization is expected to lead to an unsustainable energy demand for cooling. The 2018 report of the International Energy Agency on ‘The Future of Cooling’~\cite{IEA2018} states that without action to address energy efficiency, energy demand for space cooling will more than triple by 2050. This would consume as much electricity as all of China and India today.
Widespread use of natural ventilation, which uses the natural forces of wind and buoyancy to ventilate and cool a building, could curb this unsustainable increase in energy demand~\citep{Walker2018,Ramponi_2018, etde_290598}. 


In moderate climates, buoyancy-driven ventilation can be a very effective cooling strategy, bringing in colder outdoor air through windows, while flushing out warmer indoor air through openings at a higher height. To support this ventilation mechanism, naturally ventilated office buildings often employ atria, which connect multiple floors and provide the required height differences to support buoyancy-driven ventilation~\citep{MOOSAVI2014654, HOLFORD2003409, HUSSAIN201218}. When designing such buildings, an important challenge is the accurate prediction of the flow and cooling rates that will be achieved~\citep{HOLFORD2003409, MOOSAVI2014654, CHEN2009848}. The models have to represent the complex flow and heat transfer phenomena, and they also have to account for the effects of the highly variable building operating and weather conditions. 

A wide range of modeling approaches, ranging from building energy models to computational fluid dynamics (CFD), can be used to assess natural ventilation system performance (e.g.~\citep{CHEN2009848, ETHERIDGE201551, ZHAI20152700}).
CFD is an increasingly used modeling technique, including for the design of natural ventilation systems in buildings with atria (e.g.~\citep{JI20071158, HORAN20081577, GAN2004735, TAN20051049, Pan2010StudyOS}), since it has the advantage that it solves for the detailed flow and temperature field, eliminating the use of empirical correlations for flow and heat transfer rates. However, CFD still introduces model simplifications and assumptions in the turbulence models, geometrical simplifications, and boundary conditions. As such, it is essential to establish confidence in the predictive capabilities by performing model validation. 

To provide a realistic assessment of a model's capability to predict natural ventilation flow in full-scale operational buildings, this model validation is ideally performed using field measurements~\citep{CHEN2009848, OMRANI2017182, OMRANI20171, CACIOLO2012202, MOOSAVI2014654, BLOCKEN201469, VANHOOFF2012330}. Considering buildings with atria specifically, a few studies have pursued full-scale validation. Lau et al. investigated the vertical temperature profiles and energy savings of a 25-m tall exhibition atrium with displacement ventilation using full-scale measurements and CFD simulations~\citep{Lau03035917}. The influence of internal load and wall surface temperature were studied using CFD simulations, and the findings indicated that imposing the measured surface temperature is important for simulating displacement ventilation. Hussian and Oosthuizen used CFD to predict the thermal conditions in a building with a hybrid ventilation system~\citep{HUSSAIN2012152}. The study compared predictions obtained with four RANS turbulence models to full-scale measurements, showing good agreement. Joseph and Donal modeled air flows and temperature inside a two-story naturally ventilated office building~\citep{Joseph2005}. Comparison with full-scale measurements indicated a reasonable correlation between the numerical predictions and the measurements. Rundle et al. simulated air flows and heat transfer in the atrium space in an institutional building and compared the results to in-situ experiments ~\citep{RUNDLE20111343}. They concluded that the CFD simulation predicts the experimental data well given accurate descriptions of boundary conditions. These studies seem to confirm that CFD can provide accurate predictions of natural ventilation flow. However, they share one important limitation: in each of these studies, the computational domain only considers the interior of the buildings, and the natural ventilation flow rates are prescribed as a boundary condition at the window openings. As such, the capability of CFD to predict natural ventilation flow rates, and the resulting indoor temperatures, remains to be determined. From a design point-of-view, it is essential to investigate whether CFD can correctly predict natural ventilation flow rates given initial and boundary conditions for the temperatures and far-field wind conditions. To achieve this, the model should jointly solve for the indoor and outdoor environment. 

The objective of the present work is to address this gap in full-scale model validation of buoyancy-driven natural ventilation in atrium buildings. First, we will validate the predictive capabilities of transient CFD simulations for buoyancy-driven natural ventilation against full-scale experiments in an operational atrium building. Second, we will quantify the sensitivity of the CFD results to the thermal boundary conditions that are specified in the simulations. The simulations calculate the natural ventilation flow rate as part of the solution, i.e., by solving for the coupled indoor and outdoor velocity and temperature fields. The focus is on validating predictions of the indoor air temperature field against the full-scale measurements, considering the volume-averaged indoor air temperature, the temperature in different zones, and the temperature ranges on each floor. Full-scale measurements of the outdoor air and thermal mass temperatures are used to define the thermal boundary conditions in the simulations. Uncertainty quantification (UQ) is used to represent uncertainties in these measured initial and thermal boundary conditions. Propagation of these uncertainties through the CFD predictions generates predictions with confidence intervals, supporting more meaningful comparison to the field measurements, as well as the sensitivity analysis. Specifically, a variance-based sensitivity analysis will enable identifying the dominant uncertain parameters defining the thermal boundary and operating conditions. 
These results will complement previous sensitivity studies\mbox{~\citep{GILANI2016299, Castillo}} investigating the effects of turbulence modeling and discretization and solution methods.

The remainder of the paper is organized as follows. First, the descriptions of the full-scale experiment and the target building are given in Section~\ref{section:field_measurements}, followed by the description of the CFD model in Section~\ref{section:CFDmodel}. Section~\ref{section:results} discusses the results, which are divided into three subsections: Section \ref{subsection:singlecase} presents the validation of CFD predictions for a deterministic case, Section~\ref{subsec_results_cfd_uq} presents the validation of the CFD predictions with UQ and the results of the sensitivity analysis. Lastly, concluding remarks are given in Section \ref{Conclusions}.

\begin{figure}[H]
\includegraphics[width=\linewidth]{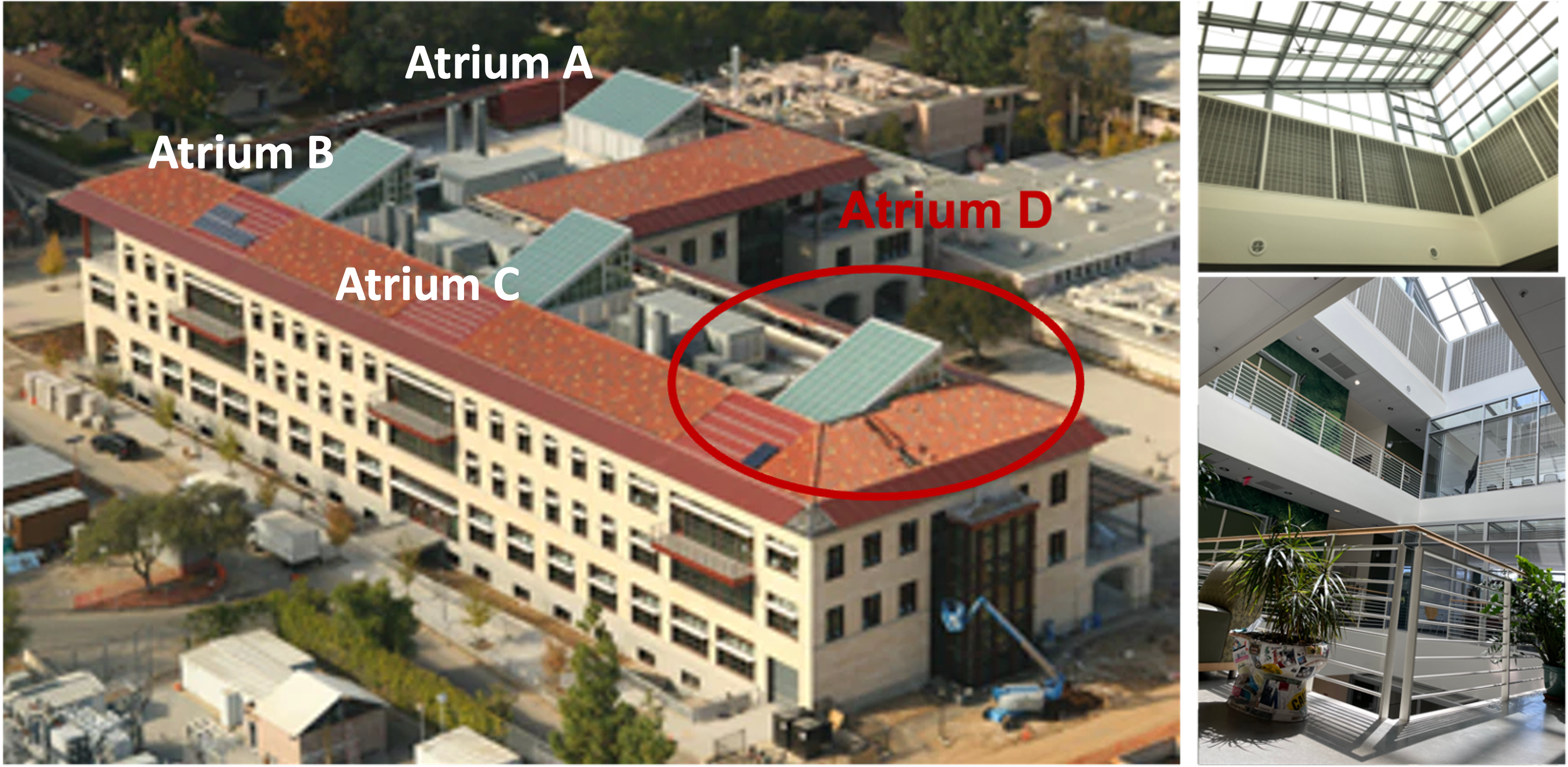}
\caption{Y2E2 building, indicating atrium D (left); louvers at the top of Atrium D (top right); indoor view of Atrium D (bottom right)~\citep{CHEN2021}}
\label{fig:atriumD}
\end{figure}

\section{Overview of building geometry and measurement data}
\label{section:field_measurements}
In this section, we first provide a description of the building geometry and the buoyancy-driven night-time natural ventilation system. Subsequently, we introduce the measurements, including the locations of the sensors in the wireless temperature sensor network. Finally, we present the experimental data that will be used for validation, considering a night during which buoyancy was the dominant driving force for the natural ventilation process. 

\subsection{Description of the Y2E2 building}
\label{subsection:Y2E2}
The Yang and Yamazaki Environment and Energy (Y2E2) Building is located on the Science and Engineering Quad on Stanford University's campus in California (Fig.~\ref{fig:atriumD}). The building has four 24.4 m atria, spanning one basement level and three above-ground stories, to support night-time passive cooling of the common spaces (hallways, open areas, and lounges connected to the central atria). The building is equipped with mechanically operated windows on each floor and with motorized louver banks on top of each atrium. During the natural ventilation process, cool air enters through the open windows, while warm air leaves through two open louver banks on the leeward side of the atrium. The cold airflow cools the exposed concrete floors of the building so that it can absorb heat loads during the following day. The windows and louver banks are closed when the indoor air temperature drops below a temperature set point.

\subsection{Description of the measurement set-up}
\label{subsection:experiment}
 
Measurements were performed during eight nights from 7:30 p.m. to 6:00 a.m. (the next day). On all nights, the outdoor temperature was lower than the volume-averaged indoor air temperature at the start of the experiment\mbox{~\citep{chen_2022_exp}}. For this initial validation study, we consider one of the nights during which the natural ventilation was driven only by buoyancy, i.e. wind effects were negligible. The first half hour was used to measure the initial conditions with the windows and louver banks closed. At 8:00 p.m. the mechanically operated windows on each floor and the eastward and northward motorized louver banks were opened simultaneously; they were closed again at 6:00 a.m. the next day. During the experimental night, the doors connecting Atrium A\&B and Atrium C\&D were closed. In combination with the symmetric geometry of Atrium C\&D, this allowed us to focus the full-scale experiment and the CFD simulation on Atrium D only.  
 
\begin{figure}[H]
\includegraphics[width=\linewidth]{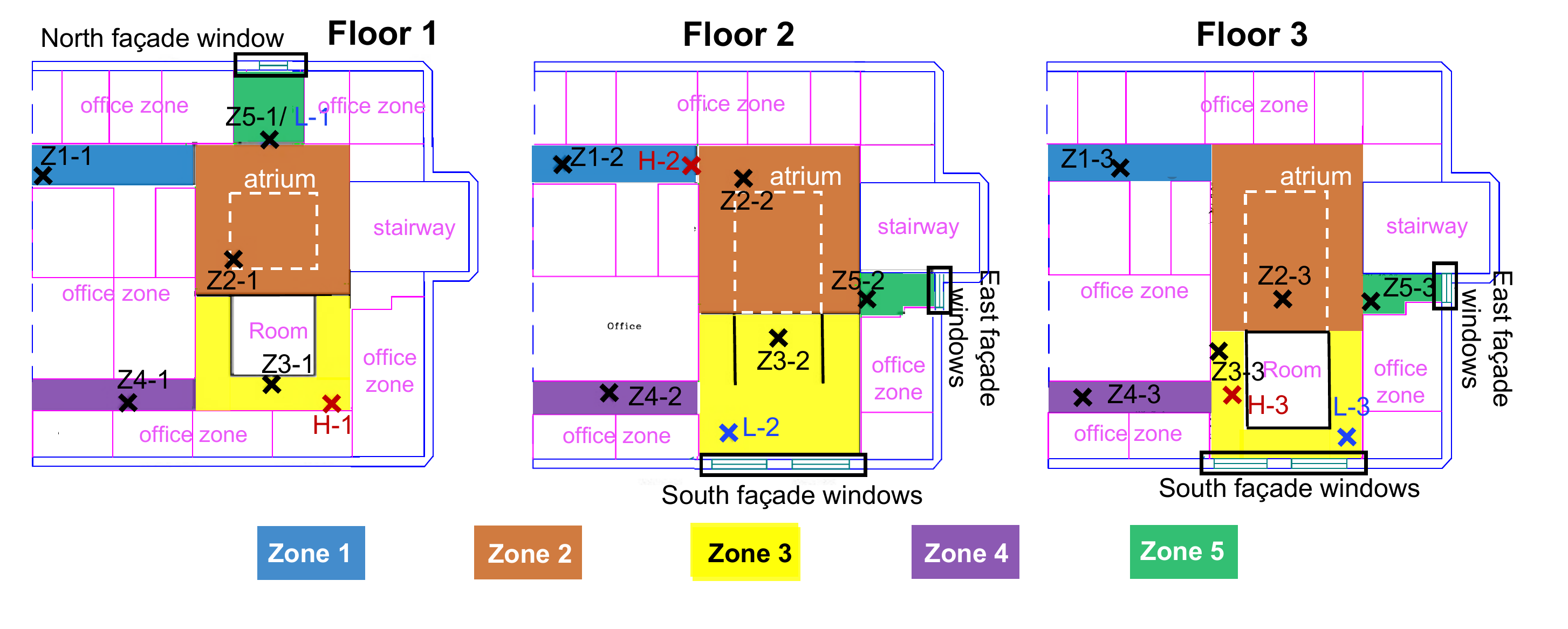}
\caption{Map of the zones, indicating the temperature sensor locations for measuring average, lower, and higher temperatures using black, red, and blue crosses, respectively.~\citep{chen_2022_exp}}
\label{fig:sensor_location}
\end{figure}

Temperatures were measured at 20 locations using a wireless temperature sensor network, as shown in Fig. \ref{fig:sensor_location}. Each location has one thermistor to measure the indoor air temperature, and one to three additional thermistors to measure nearby floor, wall, and ceiling temperatures (see Table~\ref{tab:list_sensors}). The temperature sensors were calibrated to have an accuracy of $\pm$0.3 \textdegree{}C. The outdoor temperatures ($T_{out}$) are recorded by a building sensor located on the rooftop, while the nearby  Stanford Weather Station provides information on the wind speed and direction. The indoor air and building surface temperatures were recorded every second, while the outdoor temperatures were recorded every minute. 

\begin{table}[H]
\begin{centering}
\begin{tabular}{lccc}
\hline
                       & Sensor  & Height {[}m{]} & Number of sensors          \\ \hline
\multicolumn{1}{c}{}   & Z1-1      & 2              & 3 (Air/Floor/Wall)         \\
                       & Z2-1      & 1.5            & 2 (Air/Floor)              \\
Floor 1                & Z3-1      & 2              & 2 (Air/Floor)              \\
                       & Z4-1      & 2              & 3 (Air/Floor/Wall)         \\
                       & Z5-1(L-1) & 0.5            & 2 (Air/Floor)              \\
                       & H-1       & 3              & 4 (Air/Floor/Wall/Ceiling) \\ \hline
\multicolumn{1}{c}{} & Z1-2      & 2              & 3 (Air/Floor/Wall)         \\
                       & Z2-2      & 2              & 2 (Air/Floor)              \\
                       & Z3-2      & 1.5            & 2 (Air/Floor)              \\
Floor 2                & Z4-2      & 2              & 3 (Air/Floor/Wall)         \\
                       & Z5-2      & 0.5            & 3 (Air/Floor/Wall)         \\
                       & H-2       & 2.5            & 4 (Air/Floor/Wall/Ceiling) \\
                       & L-2       & 0.5            & 2 (Air/Floor)              \\ \hline
\multicolumn{1}{c}{}   & Z1-3      & 2              & 3 (Air/Floor/Wall)         \\
                       & Z2-3      & 1.5            & 2 (Air/Floor)              \\
                       & Z3-3      & 1              & 2 (Air/Floor)              \\
Floor 3                & Z4-3      & 2              & 3 (Air/Floor/Wall)         \\
                       & Z5-3      & 1              & 3 (Air/Floor/Wall)         \\
                       & H-3       & 2.5            & 4 (Air/Floor/Wall/Ceiling) \\
                       & L-3       & 0.5            & 3 (Air/Floor/Wall)         \\ \hline
\end{tabular}
\caption{List of temperature sensors with heights from floor level in Atrium D~\citep{chen_2022_exp}}
\label{tab:list_sensors}
\end{centering}
\end{table}

The locations for the temperature sensors were selected based on a CFD- and UQ-based design of experiments presented in Chen and Gorlé~\citep{CHEN2021}. The objective of this design of experiments was to identify sensor locations that provide an accurate estimate of the volume-averaged indoor air temperature and the indoor air temperature range on each floor, given variable initial and thermal boundary conditions. To reflect the significant spatial variability in the air temperature during the natural ventilation process, each floor is divided into five zones, as shown in Fig. \mbox{\ref{fig:sensor_location}}:
\setlist{nolistsep}
\begin{itemize}[noitemsep]
     \item Zones 1 and 4: two hallway zones with offices on both sides
     \item Zone 2: the zone that includes the atrium connecting the different floors
     \item Zones 3 and 5 (except Zone 3 on floor 1): two zones that are adjacent to the windows
\end{itemize}
In Fig.~\ref{fig:sensor_location}, the black crosses in each zone indicate the optimal sensor locations for recording the zone's volume-averaged air temperature; blue and red crosses represent sensor locations that will record lower and higher than average air temperatures on each floor respectively. Table \ref{tab:list_sensors} summarizes all sensors, also reporting the heights at which the air temperature sensors are deployed. 

\subsection{Summary of experimental data used for model validation and model inputs}

The raw indoor air and surface temperature measurements were post-processed with a 1-minute moving average. 
The indoor air temperature measurements will be used for validation in terms of:
\begin{itemize}
\item the zone-averaged air temperature in each zone, as measured at the sensor locations indicated by the black crosses in each zone in Fig.~\ref{fig:sensor_location},
\item the floor-averaged air temperatures, calculated as the volume-average of the temperatures recorded in all zones on the corresponding floor,
\item the temperature range on each floor, as measured at the sensor locations indicated by the blue and red crosses in Fig.~\ref{fig:sensor_location}.
\end{itemize}
The Y2E2-averaged air temperature, calculated as the average of all floor temperatures, will be used for the normalization of temperature contours. 
To define the thermal boundary conditions for the walls, the floor, and the ceiling, surface-averaged temperatures were estimated by taking an area-weighted average of the surface temperature measurements in the different zones on each floor. 

\section{CFD model}
\label{section:CFDmodel}
The CFD simulations are performed with ANSYS/Fluent version 2021 R1 and solve for the flow velocity and temperature field. The following subsections summarize the governing equations, the computational domain and mesh, the boundary and initial conditions, the discretization and solution methods, and the uncertain parameters and propagation method.

\subsection{Governing equations}
The CFD simulations solve the Unsteady Reynolds-averaged Navier-Stokes (URANS) equations to predict the time evolution of the temperature field in Atrium D.  
The Boussinesq approximation is used to represent the effect of buoyancy, neglecting density variations in the inertia term while retaining them in the buoyancy term. The resulting governing equations for the mean temperature and velocity of the air are given by:

\begin{equation}
 \frac{\partial \overline{u_i}}{\partial x_i} = 0
\end{equation}

\begin{equation}
    \frac{\partial \overline{u_i}}{\partial t} + \overline{u_j}\frac{\partial \overline{u_i}}{\partial x_j} = -\frac{1}{\rho_0}\frac{\partial \overline{p}}{\partial x_i} + \nu\frac{\partial^2 \overline{u_i}}{\partial x_j \partial x_j} - \frac{\partial}{\partial x_j}(\overline{u_i'u_j'})+ \beta(\overline{T} - T_0)g\delta_{i3}
\end{equation}

 \begin{equation}
     \frac{\partial \overline{T}}{\partial t} +  \overline{u_i}\frac{\partial \overline{T}}{\partial x_i} =  \frac{\partial}{x_k}\left(\frac{\nu}{Pr}\frac{\partial \overline{T}}{\partial x_k} \right) -\frac{\partial }{\partial x_i} \left(\overline{u_i'T'}\right) +Q_i
 \label{eqn:energy}     
 \end{equation}

\noindent where $u$, $\overline{u}$, $u'$ represent the instantaneous, the mean, and the fluctuating velocity, respectively; $p$ is the pressure; $T$ and $T'$ are the mean and the fluctuating temperature; $\rho_0$ and $T_0$ are the reference density and temperature respectively; $\nu$ is the dynamic viscosity; $\beta$ is the thermal expansion coefficient of the air, $Pr$ is the Prandtl number, $g$ and $\delta$ are the acceleration of gravity and the Kronecker delta respectively; and $Q_i$ is a volumetric source term representing the internal heat gains from occupants, lighting, and electronic devices. $Q_i$ is defined as an uncertain parameter, which is introduced in Section~\ref{subsection:UQ}. The Reynolds stress tensor is modeled by solving model transport equations for the stress components~\citep{gibson_launder_1978} and the turbulent kinetic energy. This model has been shown to provide accurate predictions of the mean temperature for buoyancy-driven flows~\citep{Zhang_2011}. The turbulent heat fluxes are modeled using a standard gradient diffusion hypothesis with a turbulent Prandtl number of 1.0.

\subsection{Computational domain and mesh}
\begin{figure}[htbp]
\includegraphics[width=\linewidth]{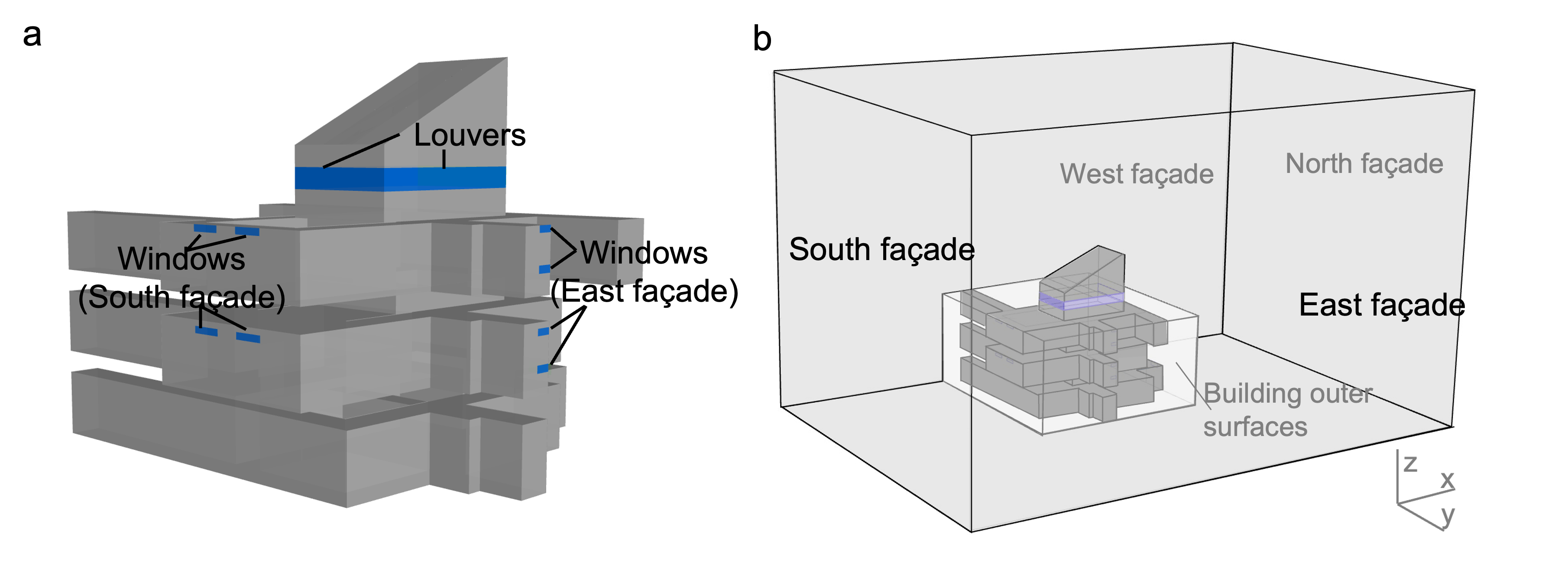}
\caption{Perspective view of (a) the geometry of Atrium D, and (b) the entire computational domain, the grad dotted line highlights the outline of Atrium D.}
\label{fig:computational_domain}
\end{figure}

Fig. \ref{fig:computational_domain} shows the building model and the computational domain. The size of the domain is 52.7 m (x-axis) by 74.9 m (y-axis) by 46.5 m (z-axis). The building geometry includes the common areas and hallways of Y2E2 (shown in Fig. \ref{fig:computational_domain}(a)). The far-field boundaries are at least 25 m (about one building height) away from the building, which is sufficiently large for considering buoyancy-driven natural ventilation. The symmetry boundary condition is imposed on the western far-field boundary cutting through the building hallways, mimicking the effect of the adjacent atrium.

\begin{figure}[htbp]
\includegraphics[width=\linewidth]{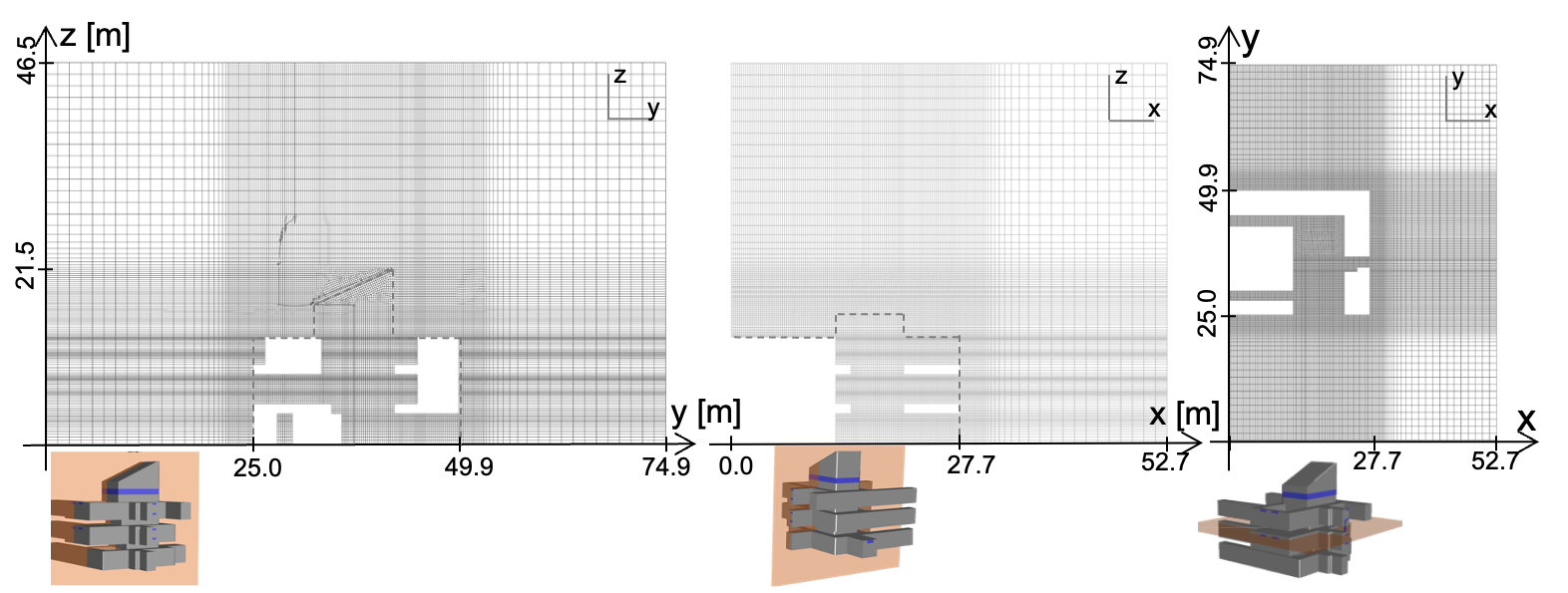}
\caption{Computational grid on a two vertical planes (left and center), and a horizontal plane (right).}
\label{fig:computational_grid}
\end{figure}

Fig. \ref{fig:computational_grid} illustrates the computational grid, which consists of 2.6 million hexahedral cells and was generated using the ANSYS Meshing tool. Inside the building, the cell size is less than 0.25 m in all locations. Gradual grid refinement is introduced near the windows and louvers (marked by the blue surfaces in Fig. \ref{fig:computational_domain}(a)) with the smallest cell size 0.09 m. The grid is gradually coarsened away from the building, with a cell expansion ratio of 1.3~\citep{franke_2004}. Near the wall, the y+ values range from 30 to 200, and equilibrium wall functions are used~\citep{Launder_1972}.

A mesh sensitivity study was carried out using a finer mesh with 8.5 million cells. Since the quantity of interest is the temperature field in the building, the mesh sensitivity study compared both zone-averaged air temperatures and point-wise air temperatures at the field measurement sensor locations. The differences between the results obtained with the two meshes were less than 0.12 \textdegree{}C throughout the entire simulation period. This negligibly small difference supports using the coarser mesh for the results presented in this paper.

\subsection{Initial and boundary conditions}
On the southern, eastern, northern, and top far-field boundaries, a zero gradient boundary condition is imposed for the velocity, together with the time evolution of the outdoor temperature and a constant atmospheric pressure. A no-slip adiabatic boundary condition is applied to the building's outer surfaces. On the interior walls (floors, walls, and ceilings), the measured time evolution of the floor-averaged surface temperatures are imposed. A porous jump boundary condition is applied on the window and louver openings to simulate the pressure drops introduced by their respective opening angles. These pressure drops are calculated as follows:
\begin{equation}
    \Delta p = \frac{1}{2}C\rho v^2
\end{equation}
where $\rho$ and $v$ are the air density and the velocity orthogonal to the openings respectively; the constant $C$ is equal to 1.65 for the window openings and 2.00 for the louver openings; these values were determined from CFD simulations of the flow through these specific opening geometries~\citep{Hult_2011}.

The experimental data indicated limited spatial variability in the temperature field on each floor at the start of the night-time ventilation, with temperature differences less than 0.15 \textdegree{}C. Therefore, the simulations are initialized using a spatially uniform temperature for each floor, equal to the floor-averaged air temperature measured on that floor before the start of the night-time natural ventilation. The initial outdoor temperature is equal to the building's outdoor sensor temperature measurement at that time. The temperature measurements used to define the initial and thermal boundary conditions, as well as the estimates of the internal heat gains, have associated uncertainty, as introduced in Section~\ref{subsection:UQ}.

\subsection{Uncertainty in initial and boundary conditions and propagation method}
\label{subsection:UQ}

To quantify the effect of uncertainty in the boundary and initial conditions, three uncertain parameters are defined: the outdoor temperature, the thermal mass surface temperature, and the internal heat gains. The outdoor temperature $T_{out}$ is defined as the sum of the measured temperature and a random parameter $\Delta T_{out}$, representing the uncertainty in the sensor calibration. Similarly, the thermal wall temperatures are defined as the sum of the measured temperatures and a random parameter $\Delta T_{tm}$, representing the uncertainty in the sensor calibration, as well as potential uncertainty due to spatial variability in the surface temperature field. The internal heat gains $Q_i$ are defined as the product of the peak design heat gain and a random multiplication factor $F_{qi}$~\citep{Hult_2011}. Given that the experiments are done in the evening, the factor is specified to represent loads due to partial lighting and/or occupants, varying from 0.35 to 0.65. Considering the lack of detailed information on the distribution of the random parameters, each parameter is assumed to have a uniform distribution with the minimum and maximum values reported in Table \ref{tab:UQparameter}.

The quantity of interest (QoI) in the analysis is the time-varying indoor air temperature field during the first three hours of natural ventilation. The effects of the uncertain parameters on the QoI are quantified using a non-intrusive polynomial chaos expansion (PCE) approach since PCE provides efficient uncertainty quantification for the analysis of nonlinear dynamic systems with a low dimensional parameter space. The UQ study was performed using Dakota v6.6~\citep{Dakota_2009}. Given the assumption of uniform distributions for the uncertain parameters, Legendre Polynomials were used as the PCE basis. An isotropic tensor grid with three points for each uncertain parameter was generated based on the Gauss-Legendre rule, leading to a total of 27 CFD simulations. The deterministic polynomial coefficients were calculated using the Legendre quadrature rule. The resulting predictions for the mean and the 95\% confidence interval of the time evolution of the indoor air temperatures are determined from probability distributions based on 10,000 samples of the PCE representations, shown in Section \ref{section:results}.
 
\begin{table}[H]
\begin{tabular}{ccc}
\hline
Uncertain parameter & Min  & Max  \\ \hline
$F_{qi}$ [-]      & 0.35 & 0.65 \\
$\Delta T_{tm}$ [\textdegree{}C] & -0.4    & 0.4    \\
$\Delta T_{out}$ [\textdegree{}C]  & -0.4    & 0.4    \\\hline
\end{tabular}
\caption{Uncertain parameters in the CFD simulations}
\label{tab:UQparameter}
\end{table}

\subsection{Discretization and solution method}
The PISO (Pressure-Implicit with Splitting of Operators) algorithm is used for pressure-velocity coupling. Discretization is done using second-order schemes in space and a second-order implicit scheme in time. A small time step is used at the beginning of the simulation to capture the fast initial transient in the temperature and flow field when the windows are opened. The time step is then gradually increased to reduce the computational costs as the temporal changes in the flow and temperature fields become smaller. Based on a sensitivity test with varying time steps, the optimal time steps are 0.5 s, 1.0 s, 2.0 s, and 5.0 s for 0 to 5 mins, 5 to 15 mins, 15 to 30 mins, and 30 to 60 mins, respectively. For the remaining two hours of the simulation, the time step is set to 10 s when a quasi-steady state is achieved. At each time step, the residuals of the continuity and momentum equations decrease below $10^{-4}$ and $10^{-6}$ respectively before advancing the simulation. The simulation is stopped after the first three hours; during normal operation, the night-time natural cooling is usually disabled beyond this time (i.e. windows and louvers are closed) since the indoor air temperature will drop below the temperature set point. 

\section{Results and discussion}
\label{section:results}

\subsection{Validation of baseline CFD predictions}
\label{subsection:singlecase}

This section analyzes the results of the CFD simulation with the baseline boundary and initial conditions, where $\Delta T_{out} = 0$ \textdegree{}C, $\Delta T_{tm} = 0$ \textdegree{}C, and $F_{qi} = 0.38$. Section \ref{subsubsec_contours_deterministic} provides a qualitative picture of the flow and temperature fields using contour plots. Subsequently, Section \ref{subsubsec_volavg_deterministic} and Section \ref{subsubsec_results_cfd_minmax} quantitatively compare the experimental and numerical results for the time evolution of the floor- and zone-averaged air temperatures and the temperature ranges on each floor. Lastly, the relative contributions of different uncertain parameters to the variance in the predicted zone-averaged air temperature are quantified in Section~\ref{subsection_sensitivity}.

\subsubsection{Analysis of flow and temperature fields}
\label{subsubsec_contours_deterministic}

\begin{figure}[tbp]
\includegraphics[width=\linewidth]{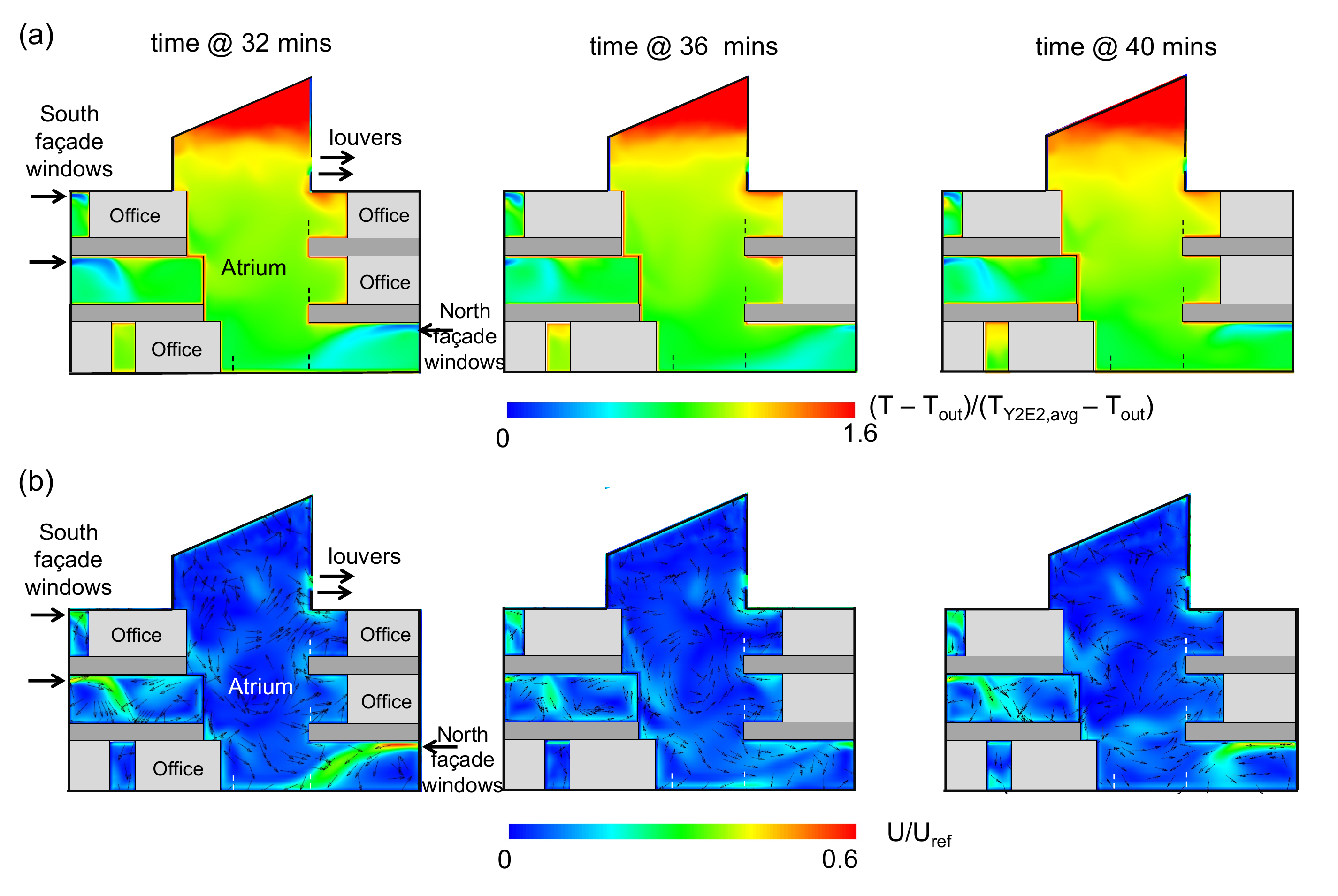}
\caption{Contours of (a) instantaneous dimensionless temperature difference and (b) instantaneous dimensionless velocity on a vertical plane at 32 minutes, 36 minutes, and 40 minutes from the start of night-time natural ventilation.}
\label{fig:contour_ver}
\end{figure}

\begin{figure}[tb]
\includegraphics[width=\linewidth]{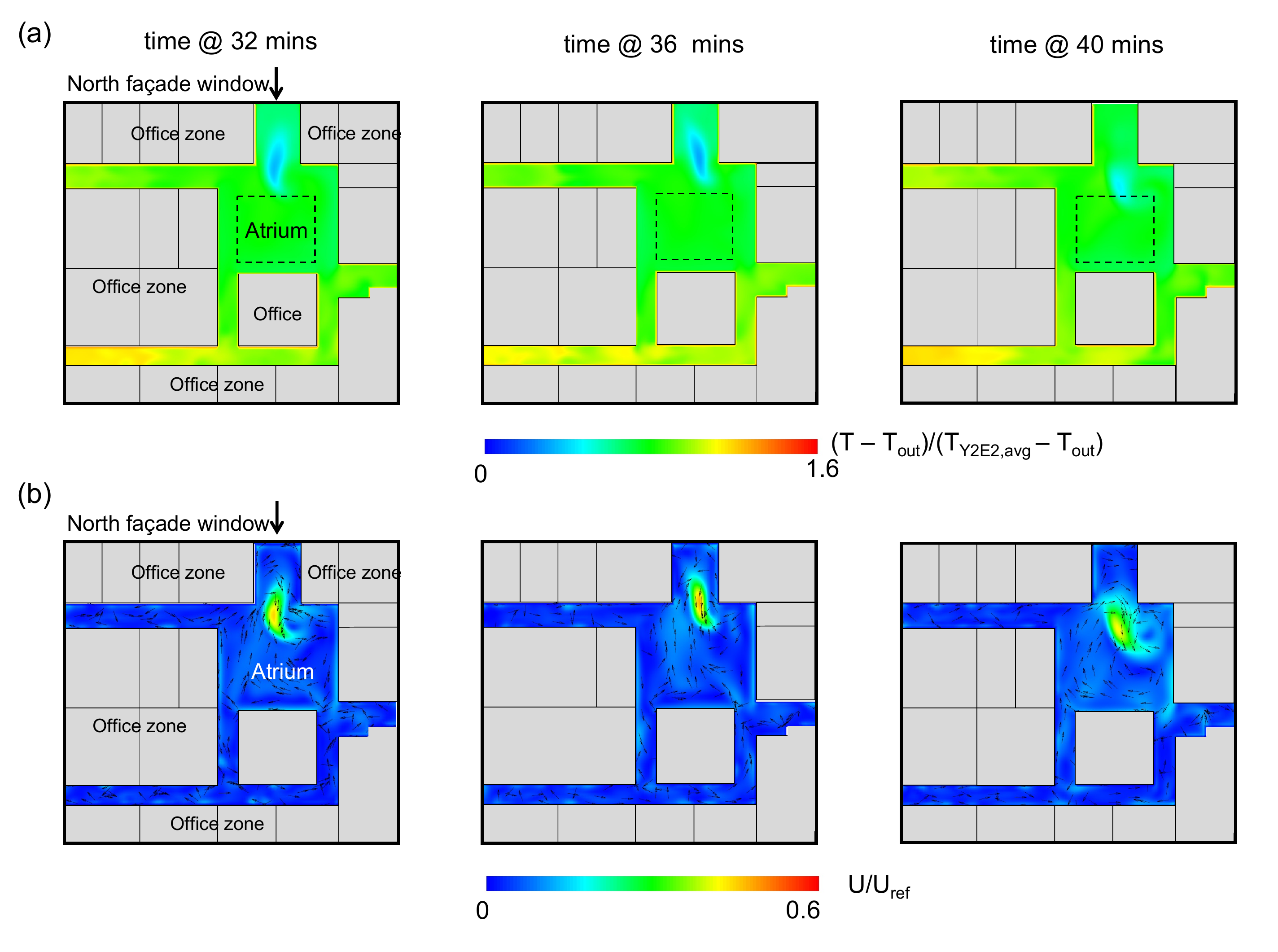}
\caption{Contours of (a) instantaneous dimensionless temperature difference and (b) instantaneous dimensionless velocity on a horizontal plane of Floor 1 (2 m above the ground) at 32 mins, 36 mins, and 40 mins from the start of night-time natural ventilation.}
\label{fig:contour_hor}
\end{figure}

Fig. \ref{fig:contour_ver} and \ref{fig:contour_hor} visualize contour plots of the instantaneous temperature and velocity magnitude inside the building predicted by the baseline CFD simulation. The plots are shown on a vertical plane that cuts through some of the south and north facade window openings and the louvers, and on a horizontal plane at 2 m height on floor 1. The figures include three time instances at a four-minute interval, taken about half an hour into the night-time natural ventilation. The temperature field is presented using the dimensionless temperature difference $(T-T_{out})/(T_{Y2E2,avg}-T_{out})$. The velocity is normalized using the buoyant reference velocity ($U_{ref}$):

\begin{equation}
    U_{ref}  = \sqrt{2g\Delta h\frac{T - T_{out}}{T_{out}}},
\end{equation}
where $\Delta h$ represents the height difference between the louver and the window opening on floor 1. 

The contour plots on the vertical plane in Fig. \ref{fig:contour_ver} show how cool air enters the building through the north-facing window on floor 1 and the south-facing windows on floors 2 and 3. The warm air is concentrated at the top of the atrium and leaves the building through the louver banks on the north facade. Comparison of the three time instances reveals that the flow near the window openings is highly unsteady, exhibiting significant fluctuations in the velocity field. This unsteadiness is most pronounced in the contour plots of the dimensionless velocity of the inflow jet through the north-facing window on floor 1. 

Fig. \ref{fig:contour_hor} shows the contour plots on a horizontal plane just below this window, at 2 m above the ground on floor 1. The plot shows how the cool air entering from the window above forms a region of colder air that extends from Zone 5 into Zone 2. The flow in this region is confirmed to be highly unsteady, with the inflow jet oscillating in both the horizontal and vertical directions. The unsteadiness of the flow and temperature fields illustrates the importance of using unsteady models to simulate the buoyancy-driven natural ventilation process. The plot also visualizes the significant horizontal variability in the indoor temperature. Zones 2 and 5, which are close to the window, show the lowest temperatures due to the cool inflow jet. Zone 4, which is the hallway away from the window, is warmer than other zones since the flow path from the window to the louvers does not pass through this zone.

\subsubsection{Comparison of zone-averaged indoor air temperature}
\label{subsubsec_volavg_deterministic}

\begin{figure}[H]
\includegraphics[width=\linewidth]{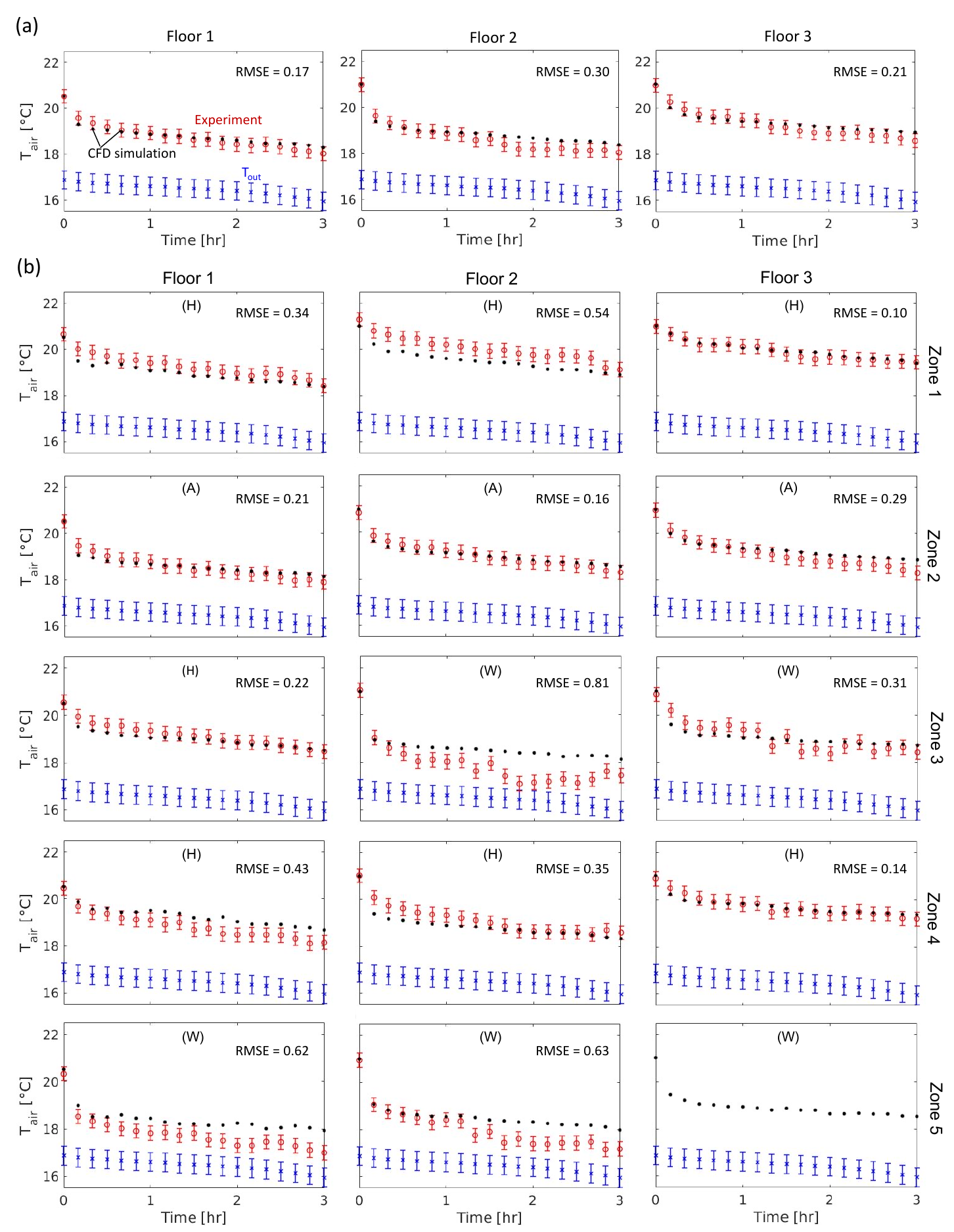}
\caption{Comparison of (a) the floor-averaged air temperature, and (b) the zone-averaged indoor air temperatures from the baseline CFD simulation and the full-scale experiment. 0 hr indicates the start of night-time natural ventilation. RMSE represents the root-mean-square errors between the CFD simulation and the measurements in the unit of \textdegree{}C. (A), (H), (W) indicate the atrium, hallway, and window zones, respectively.}
\label{fig:Tzoneavg_onecase}
\end{figure}

Fig. \ref{fig:Tzoneavg_onecase} presents the comparison of the CFD predictions and full-scale measurements of the floor- and zone-averaged air temperature, as well as the outdoor temperature measurements. The error bars on the measurements represent the measurement uncertainty of $\pm$0.3 \textdegree{}C for the floor- and zone-averaged air temperature and $\pm$0.4 \textdegree{}C for the outdoor temperature. Fig. \ref{fig:Tzoneavg_onecase}(a) indicates that the CFD simulation can accurately predict the floor-averaged temperatures on all floors, with root-mean-square errors (RMSEs) less than or equal to the measurement uncertainty. On floors 1 and 3, the predicted floor-averaged air temperature falls within the experimental uncertainty throughout the entire passive cooling period. On floor 2, the CFD predicts a slightly higher floor-averaged air temperature than the measurements for the second half of the modeled time period. 

\begin{figure}[htbp]
\includegraphics[width=\linewidth]{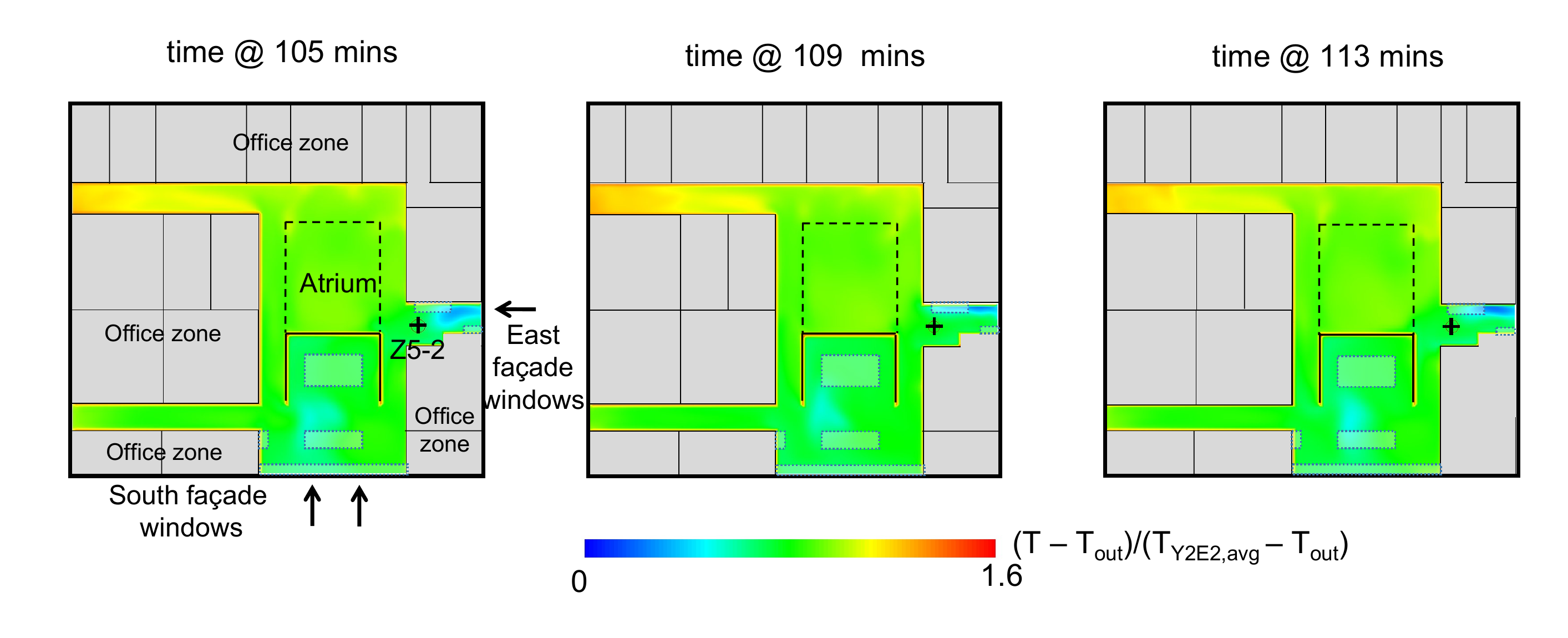}
\caption{Contours of instantaneous dimensionless temperature difference on a horizontal plane at 0.5 m from the floor level of floor 2. The outline of furniture that is not represented in the CFD model is highlighted using dotted blue boxes.}
\label{fig:temp_contour_f2f3}
\end{figure}

Figure~\ref{fig:Tzoneavg_onecase}(b) presents the comparison between the CFD predictions and measurements for the zone-averaged air temperatures. The columns present the different floors, while the rows present the different zones on each floor. The measurements from Floor 3 Zone 5 are excluded from the analysis because of a sensor failure during the experiment. The RMSEs vary from 0.1\textdegree{}C to 0.81 \textdegree{}C, indicating overall good agreement. The UQ analysis in Section \ref{subsection:UQ} will investigate to what extent the discrepancies can be explained by uncertainty in the boundary conditions or internal heat gains. However, one noteworthy observation from Figure~\ref{fig:Tzoneavg_onecase}(b) is that the largest discrepancies on each floor appear in the zones closest to window openings, i.e. in Floor 1 Zone 5, Floor 2 Zones 3 and 5, and Floor 3 Zone 3. In these regions, the discrepancies are larger during the second half of the nigh-time ventilation and the sensor measurements exhibit oscillations. Based on the flow visualizations in Figs.~\ref{fig:contour_ver} and~\ref{fig:contour_hor}, a possible explanation for these discrepancies could be a slight deviation of the location of the oscillating inflow jets in the CFD compared to reality. Such a deviation can be caused by geometrical differences between the CFD model and reality, such as the detailed window geometry being represented by a porous jump condition or the presence of furniture~\citep{Zhuang2014CFDSO}. To explore this explanation, Fig.~\ref{fig:temp_contour_f2f3} depicts additional temperature contours on floor 2, which has the largest RMSEs, around the time at which the discrepancies are larger. The plot includes the outline of some large file cabinets and desks that are not represented in the CFD model. Considering Zone 5, these cabinets likely create a deviation of the inflow jets towards the sensor location. Similarly, the furniture in Zone 3 is likely to affect the flow pattern. This result indicates that the current validation effort could be further improved by considering the presence of large furniture in both the CFD-based design of experiments and the validation simulations, or by equipping zones close to windows with multiple sensors.

\subsubsection{Comparison of temperature range}
\label{subsubsec_results_cfd_minmax}

\begin{figure}[H]
\includegraphics[width=\linewidth]{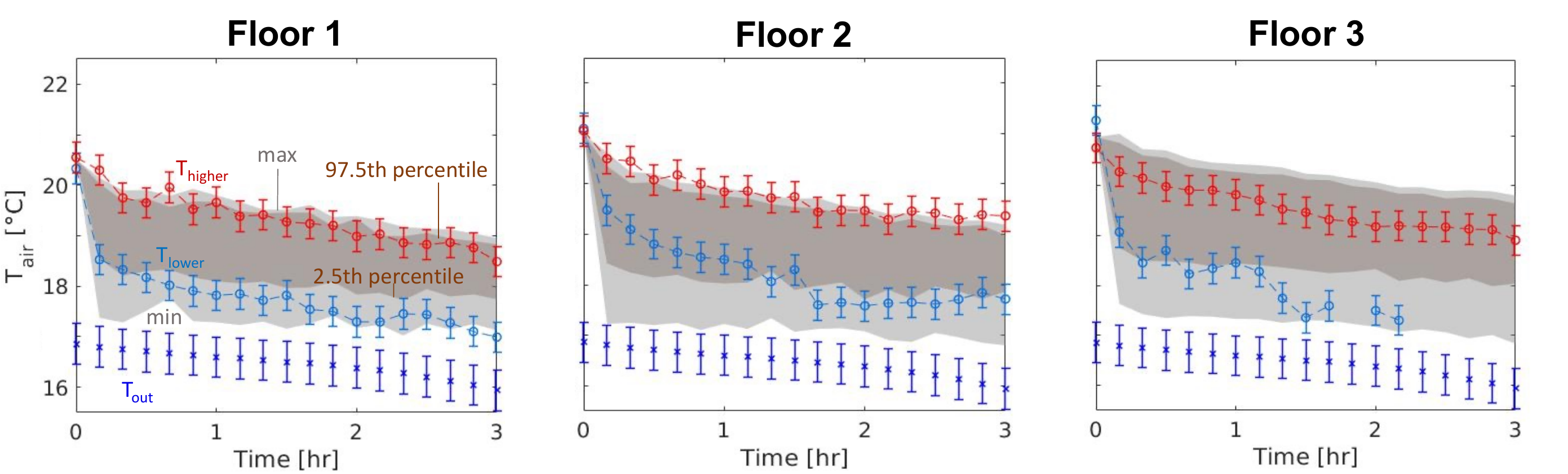}
\caption{Comparison of the temperature range from CFD simulation, including the range between the 2.5th to the 97.5th percentile of indoor temperature on a single floor(brown shaded zone), the range between the minimum temperature to the maximum temperature on a single floor(gray shaded zone), with the measured temperature range defined by measurements from the lower temperature sensor ($T_{lower}$) and the higher temperature sensor ($T_{higher}$) (see blue and red crosses in Fig. \ref{fig:sensor_location}) on each floor.}
\label{fig:Tminmax}
\end{figure}

Fig.~\ref{fig:Tminmax} presents the minimum and maximum, as well as the 2.5th and 97.5th percentile of the indoor temperature distribution predicted by the CFD. The computational results are compared to the temperatures measured at the sensors located to record lower and higher than average temperatures on each floor. Following the design of experiments, we expect these measurements to roughly corresponding to the 2.5th and 97.5th percentile of the indoor temperature distribution~\citep{CHEN2021}. 

The computational results in Fig.~\ref{fig:Tminmax} indicate that the distribution of the temperature has a short tail at the maximum temperature end, with the difference between the 97.5th percentile and the maximum value only about 0.4 \textdegree{}C. In contrast, the distribution has a long tail at the minimum temperature end, where the difference between the 2.5th percentile and the minimum temperature can be as large as 2.0 \textdegree{}C. This observation is in agreement with the flow patterns visualized in Section~\ref{subsection:singlecase}, which indicated that the colder outdoor air flows into the building through the windows in relatively localized unsteady jets with the coldest air at the core. 

Comparison of the numerical and experimental results indicates that on the first and second floor, the width of the measured temperature range agrees well with the width of the predicted 2.5th to 97.5th percentile of the indoor temperature, especially during the first hour of natural ventilation. On floor 1, the measurements from the lower temperature sensors are closer to the minimum temperature after the first hour. On floor 3, the high temperature measurement is lower than the CFD prediction for the 97.5th percentile by 0.45 \textdegree{}C. On the low temperature end, the behavior on floor 3 is similar as on the other floors, although the oscillations in the measurement indicate that the sensor might be located in one of the regions with the unsteady flow.

\subsection{Validation of CFD predictions with UQ for the temperature fields}
\label{subsec_results_cfd_uq}

In this section, the effects of uncertainty in the boundary and initial conditions are quantified using the PCE method introduced in Section~\ref{subsection:UQ}). The uncertainties, summarized in Table~\ref{tab:UQparameter}, arise from the experimental measurements used to define the outdoor air temperature and the indoor surface temperatures, as well as from the internal heat gains. In the following we first present the mean and 95\% confidence interval (CI) of the floor- and zone-averaged indoor air temperature predictions. Subsequently, we quantify the relative importance of the different uncertain parameters.

\subsubsection{Prediction of the mean and 95\%CI of zone-averaged indoor air temperature}
\label{subsubsec_results_cfd_uq_volavg}

\begin{figure}[H]
\includegraphics[width=\linewidth]{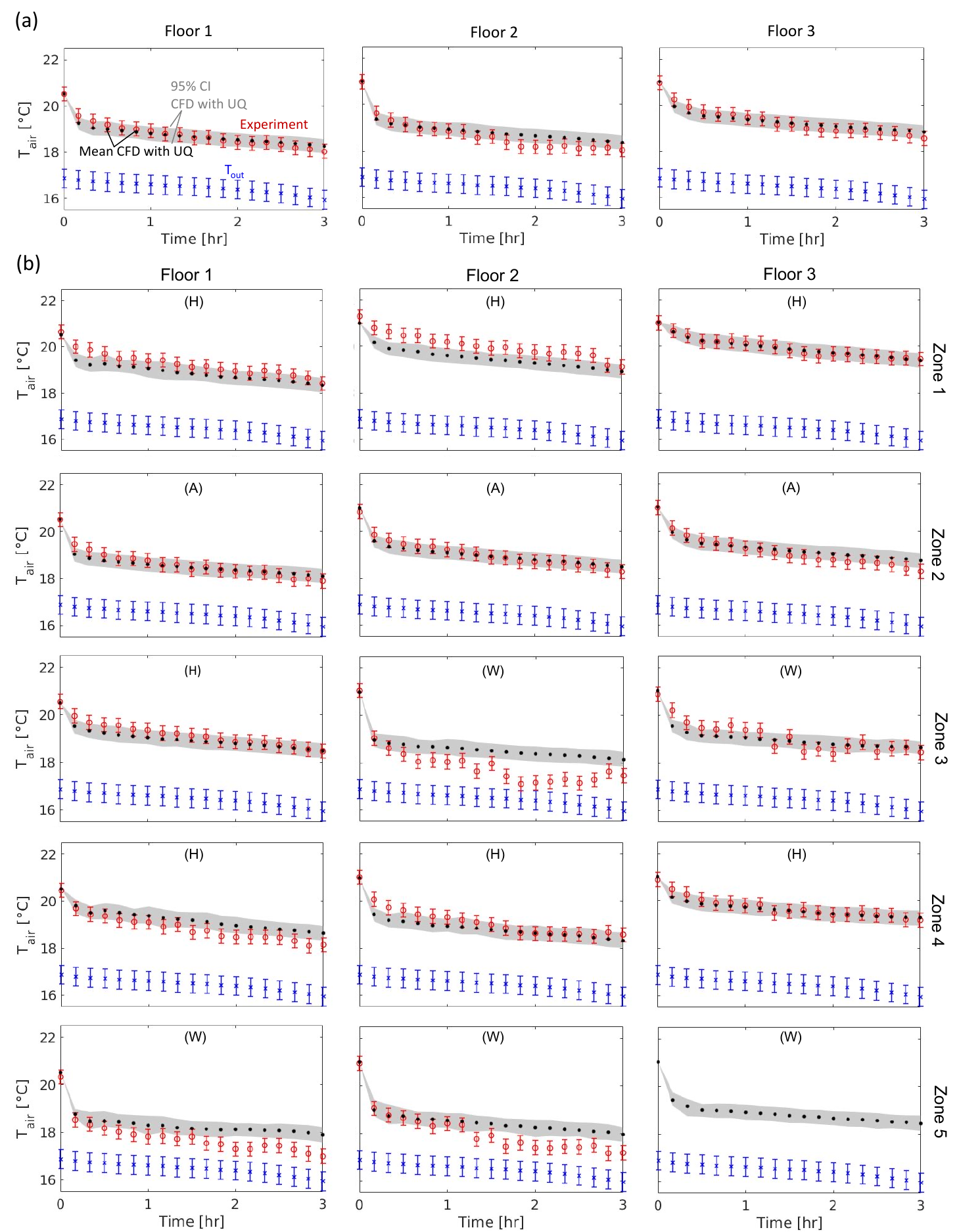}
\caption{Comparison of (a) the floor-averaged air temperature, and (b) the zone-averaged air temperatures from CFD simulations with UQ and full-scale experiment. (A), (H), (W) indicate the atrium, hallway, and window zones, respectively.}
\label{fig:Tzoneavg_uq}
\end{figure}

Fig. \ref{fig:Tzoneavg_uq}(a) compares the CFD predictions for the mean and 95\% CI of the floor-averaged air temperature to the full-scale measurements. The CFD 95\% CI overlap with the measurement error bars at all times, indicating an accurate prediction for the floor-averaged air temperature by the CFD simulations with UQ. Fig. \ref{fig:Tzoneavg_uq}(b) presents the same comparison for the zone-averaged air temperatures. As in Fig.~\ref{fig:Tzoneavg_onecase}, the columns represent the different floors, while the rows represent the different zones. Two trends can be observed. In the hallway and atrium zones (Zones 1, 2, and 4 on all floors, and Zone 3 on floor 1), the CFD 95\% CI overlap with the measurement error bars from 78\% to 100\% of the time. The maximum discrepancy between the confidence interval bounds and the measurement data is very small at 0.15 \textdegree{}C. On the other hand, considering the zones adjacent to the windows on floor 1 (Zone 3) and Floor 2 (Zone 3 and 5), the CFD 95\%CI only overlap with the measurement error bars  39\% to 61\% of the time and the maximum discrepancies are higher, up to 0.70 \textdegree{}C. Zone 3 on floor 3, which is also close to a window, is an exception with an overlap rate of 94\% between the CFD 95\% CI and the measurement error bars. As discussed in Section \ref{subsection:singlecase}, the zones close to the windows have additional uncertainty due to geometrical simplifications in the CFD model. This uncertainty could be reduced by including large furniture and a more detailed representation of the window geometry in the simulations. Overall, the effect of the geometrical simplifications is highly localized in regions close to the windows. As a result, the UQ analysis does produce accurate predictions for the 95\% CI of the floor-averaged temperatures.

\subsubsection{Sensitivity analysis}
\label{subsection_sensitivity}

This section discusses the relative contributions of the different uncertain parameters $X$ = [$F_{qi}$, $\Delta T_{tm}$, $\Delta T_{out}$] to the variance in predictions of the temperature ($T$). These relative contributions are quantified using a variance decomposition technique for global sensitivity analysis, calculating the Sobol indices ($S_{T_i}$)~\citep{SOBOL2001271}:
\begin{equation}
    S_{T_i} = \frac{E(Var(T|X_{\sim i}))}{Var(T)}
\end{equation}

\noindent where $Var(T|X_{\sim i})$ is the variance of $T$ conditioned on all uncertain parameters except $X_i$. Hence, $S_{T_i}$ quantifies the total contribution of $X_i$, including its interactions with other uncertain parameters, to the variance of $T$.

\begin{figure}[H]
\includegraphics[width=\linewidth]{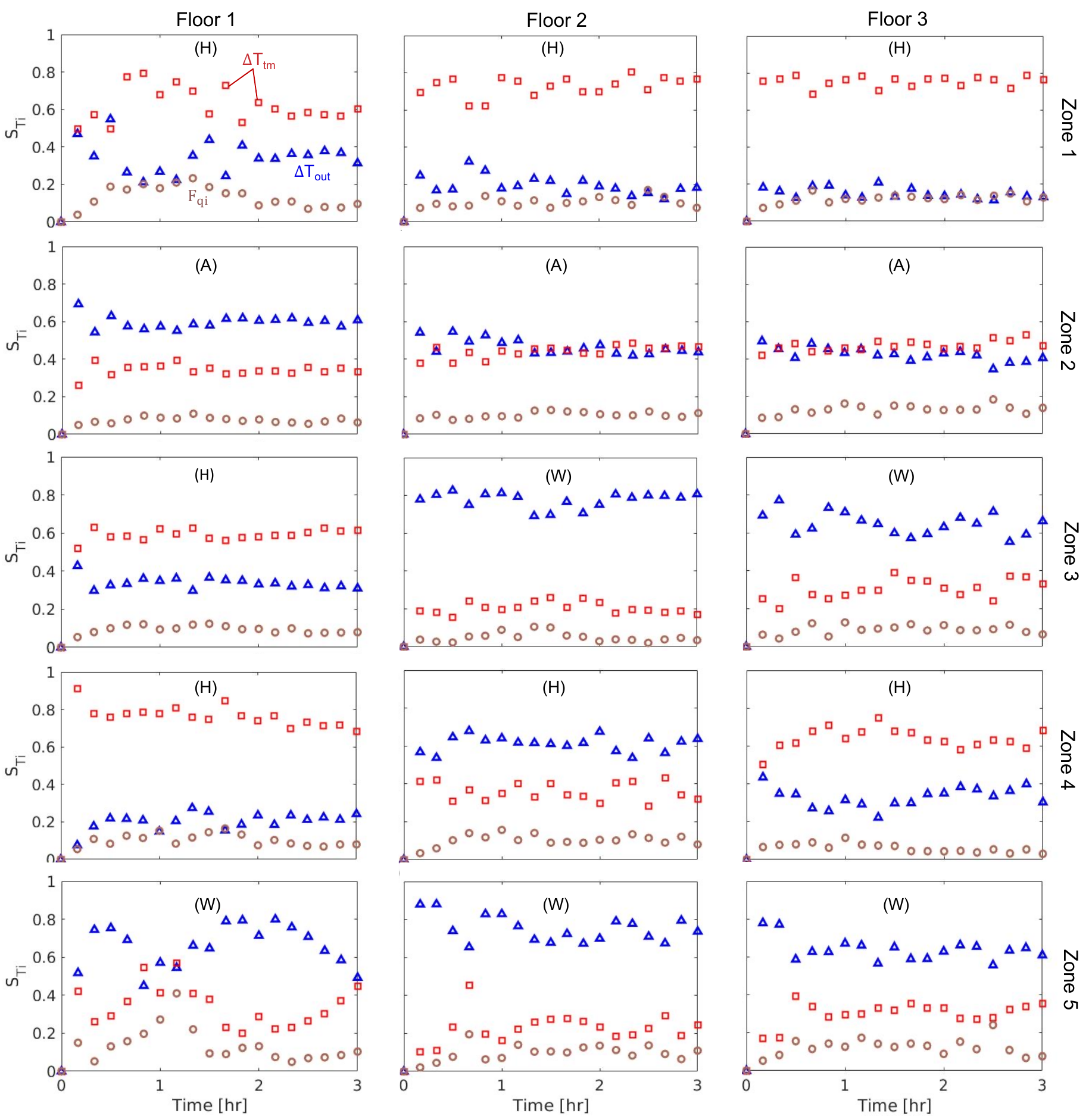}
\caption{Sobol sensitivity indices for the zone-averaged indoor air temperature. (A), (H), (W) indicate the atrium, hallway, and window zones, respectively.}
\label{fig:sTi_uq}
\end{figure}

Fig.~\ref{fig:sTi_uq} presents the Sobol indices for the zone-averaged air temperatures for each of the three uncertain parameters, with the columns representing the different floors and the rows representing the different zones. As before, the plots indicate different trends in different regions. For zones away from windows, where the ratio of the wall surface area to zone's volume is large (Floor 1 Zone 4, Floor 1 Zone 3, Floor 2 Zone 1, and Floor 3 Zone 1), the uncertainty in the thermal mass temperature $\Delta T_{tm}$ is the dominant parameter over the entire natural ventilation period. Since these zones are not directly exposed to the cool air flowing in through the window and out through the louvers, $\Delta T_{out}$ contributes less to the variance in hte temperature predictions. In contrast, in the zones adjacent to windows (Floor 2 Zone 3, Floor 3 Zone 3, and Zone 5 on all floors), $\Delta T_{out}$ is the dominant uncertain parameter. The results for the atrium zones (Zone 2 on all floors), indicate that $\Delta T_{out}$ and $\Delta T_{tm}$ have a similar impact on the zone-averaged air temperature. One deviation from these trends is that the temperature in Floor 2 Zone 4 is most effected by $\Delta T_{out}$, even though this is a hallway zone with offices on both sides. The dominant effect of  $\Delta T_{out}$ in this zone can be explained by the large open space in front of the south-facing windows in the adjacent Zone 3. This geometry induces a different flow pattern compared to Floor 3, with more outdoor air flowing into the hallway. Lastly, it is noted that $F{qi}$ has the smallest effect among the uncertain parameters. This is partially due to the fact that the natural ventilation process occurs in the evening when relatively few occupants are in the building and the usage of lights and electronic devices is significantly reduced. 

\section{Conclusions}
\label{Conclusions}

This study has demonstrated the predictive capabilities of unsteady RANS simulations with a Reynolds stress transport model for buoyancy-driven natural ventilation in an atrium building. The simulations solve for the coupled indoor and outdoor environment to produce indoor air temperature predictions that compare favorably to full-scale measurements. In terms of the floor-averaged air temperatures, the predictions and measurements agree within 0.3 \textdegree{}C, which corresponds to the measurement sensor accuracy. When considering the temperature averaged over smaller zones on each floor, a similar level of agreement is reached in zones that are not adjacent to windows. Furthermore, a UQ analysis indicates that the small discrepancies observed can be attributed to uncertainties in CFD model inputs for the thermal boundary conditions and internal heat gains. In zones adjacent to windows, the discrepancies increase to a maximum of 0.70 \textdegree{}C, and the uncertainty in the input parameters no longer fully explains these differences. Visualization of the flow pattern in these regions indicates that the inflow of cold outdoor air through the windows is an unsteady process with significant oscillations in the inflow direction. This oscillating inflow direction likely differs in the numerical predictions compared to reality due to geometrical simplifications in the CFD model. Improving the local temperature predictions would therefore require a higher-fidelity representation of the geometry; however, these larger discrepancies are due to highly localized phenomena, which do not affect the ability of the CFD model to predict floor-averaged air temperatures.

A sensitivity analysis to identify the relative importance of the different uncertain input parameters demonstrates that the boundary conditions for the thermal mass surface temperature and the outdoor temperature are dominant parameters, while the internal heat gains have a smaller effect. Zones adjacent to windows are primarily affected by the outdoor temperature, while hallway zones surrounded by offices and further away from windows are primarily affected by the thermal mass surface temperature. In the atrium space, both parameters equally influence the results. The sensitivity analysis indicates that accurate information of the thermal mass surface temperature and the outdoor temperature is key for an accurate simulation of buoyancy-driven natural ventilation. In absence of surface temperature measurements, this implies that the CFD model should be coupled with a dynamic thermal model for the building thermal mass. The development of computationally efficient strategies for such coupled models, as well as their validation, is a topic of ongoing research.

\section*{Acknowledgements}
This research was supported by a Seed Research Grant from the Center for Integrated Facility Engineering at Stanford University. 

\newpage
\bibliographystyle{unsrt}

\bibliography{ms}

\end{document}